# Four Tensors Determining Thermal and Electric Conductivities of Degenerate Electrons in Magnetized Plasma

G. S. Bisnovatyi-Kogan[a,b,]* and M. V. Glushikhina[a,]**

[a] *Space Research Institute, Russian Academy of Sciences, Moscow, 117997 Russia*
[b] *National Research Nuclear University MEPhI, Moscow, 115409 Russia*
*e-mail: gkogan@iki.rssi.ru
**e-mail: m.glushikhina@iki.rssi.ru

**Abstract**—A solution to the Boltzmann equation is obtained for a magnetized plasma with strongly degenerate nonrelativistic electrons and nondegenerate nuclei. The components of the diffusion, thermal diffusion, and diffusion thermoeffect tensors in a nonquantizing magnetic field are calculated in the Lorentz approximation without allowance for electron–electron collisions, which is asymptotically accurate for plasma with strongly degenerate electrons. Asymptotically accurate analytical expressions for the electron diffusion, thermal diffusion, and diffusion thermoeffect tensors in the presence of a magnetic field are obtained for the first time. The expressions reveal a considerably more complicated dependence on magnetic field than analogous dependences derived in the previous publications on this subject.

## 1. INTRODUCTION

The kinetic coefficients in the crusts of neutron stars and the cores of white dwarfs play an important role in the evolution of these stars. The heat fluxes and current densities are determined by heat conduction, diffusion, thermal diffusion, and the diffusion thermoeffect. To calculate these coefficients, it is necessary to know the transport properties of a dense stellar matter, in which electrons are strongly degenerate and form a nearly ideal Fermi gas, while ions are nondegenerate and form either a Coulomb liquid or a Coulomb crystal. Under such conditions, electrons are the main carriers of charge and heat and the kinetic coefficients are mainly determined by electron scattering from ions. Knowing the distributions of heat and current, one can calculate the magnetothermal evolution of a neutron star with the crust that forms a Coulomb crystal [1]. The magnetic field confines electrons in the direction perpendicular to magnetic field lines. Since electrons are the main carries of heat and current, heat and charge transport in this direction is suppressed, remaining unaffected along the field lines. The kinetic coefficients of degenerate electrons in neutron stars and white dwarfs in a magnetic field were analyzed in [2, 3].

The ratio between the electric conductivities along and perpendicular to the magnetic field was phenomenologically obtained in [2]. It is given by

$$\frac{\sigma_\perp}{\sigma_\parallel} = \frac{1}{1 + (\omega\tau)^2} \quad (1)$$

and was used in [3] to calculate the kinetic coefficients of a degenerate electron gas in the crusts of neutron stars. Here, $\omega = eB/(m_e c)$ is the cyclotron electron frequency, $\tau$ is the time between electron collisions, $e$ is the electron charge, $m_e$ is the electron mass, $B$ is the magnetic field strength, and $c$ is the speed of light. The influence of the magnetic field on the electron thermal conductivity and electric conductivity in form (1) was used in many subsequent papers. Methods of kinetic theory of gases were developed by Maxwell, Boltzmann, Hilbert, Enskog, and Chapman. These methods were described in the monograph by Chapman and Cowling [4]. They are based on solving the Boltzmann equation by the method of successive iterations. A thermodynamically equilibrium distribution function (the Maxwellian distribution for a nondegenerate gas or the Fermi distribution for a degenerate gas) is used as a zero-order approximation. The equation has no exact analytic solution in the first approximation and is solved by expansion in the Sonine (Laguerre) polynomials. The simplest, from the standpoint of calculations, is the case in which the plasma electron gas can be considered to be Lorentzian. Due to the large nucleus-to-electron mass ratio, the absolute value of the electron momentum can be assumed to be preserved in an electron–nucleus collision. In this approximation, electron–electron colli-sions are also neglected, which allows one to solve the equation for the electron distribution function in the first-order approximation analytically without using an expansion in the Sonine polynomials.

The application of the Boltzmann equation to calculate the kinetic coefficients in a gas of charged particles in the presence of a magnetic field was described in [4]. Due to the divergence of the collision integral at large values of impact parameters for particles exhibiting Coulomb interaction, the Debye length (the distance at which the particle charge is screened) should be used as the upper limit of integration with respect to the impact parameter [5, 6]. The viscosity, thermal conductivity, and diffusion coefficient, which have tensor form under the conditions of magnetic-field-induced anisotropy, were calculated for gases consisting of charged particles.

The kinetic coefficients in a nondegenerate plasma in the presence and absence of a magnetic field were calculated more accurately in [5–9] by using the Chapman–Enskog expansion method. The kinetic coefficients for fully ionized plasma of complex composition were calculated in [10]. The electric conductivity of hot plasma consisting of electrons and positively charged ions in the presence of a uniform magnetic field was numerically calculated in [11] by applying the Chapman–Enskog method and retaining 50 terms in the Sonine polynomial expansion.

The kinetic coefficients of the electric and thermal conductivities in the crusts of magnetized neutron stars with allowance for electron–electron scattering, as well as for scattering from defects of crystalline lattice and impurities, were calculated in [12]. The electric and thermal conductivities in a degenerate relativistic stellar matter were calculated in [13] by solving the Boltzmann equation in the relaxation-time approximation.

The electron thermal conductivity tensor in a dense magnetized matter was calculated in [14] by solving the Boltzmann equation by the Chapman–Enskog method for nondegenerate electrons and in the Lorentz approximation for strongly degenerate nonrelativistic electrons. In the present work, we calculate the diffusion, thermal diffusion, and diffusion thermoeffect tensors for strongly degenerate electrons in the Lorentz approximation. Asymptotically accurate analytic expressions for these kinetic coefficients in the presence of a magnetic field are obtained for the first time. The expressions reveal a substantially more complicated dependence on magnetic field than those obtained in earlier publications on this subject. The use of the kinetic coefficients calculated in this paper makes it possible to take into account the processes occurring in the crust of a neutron star more accurately. The obtained expressions can also be used to describe the transport coefficients in other magnetized objects containing free degenerate electrons.

## 2. BOLTZMANN AND TRANSPORT EQUATIONS

Let us consider electron gas in a crystalline lattice consisting of heavy nuclei and take into consideration the interaction of electrons with nondegenerate nuclei and with one another. The nuclear component of matter in the crust is in a crystalline phase. Therefore, the isotropic part of the distribution function $f_{N0}$ can differ from the Maxwellian distribution. If the mass $m_N$ of a nucleus is much larger than the electron mass $m_e$, the details of the distribution function $f_{N0}$ to within terms of $\sim m_e/m_N$ are unimportant and the calculations can be performed for an arbitrary function $f_{N0}$.

The Boltzmann equation describing time variations in the distribution function $f$ of nonrelativistic electrons in the presence of electric and magnetic fields can be written in the form [8, 9]

$$\frac{\partial f}{\partial t} + c_i \frac{\partial f}{\partial r_i} - \left[\frac{e}{m_e}\left(E_i + \frac{1}{c}\varepsilon_{ikl}c_k B_l\right) - F_i\right]\frac{\partial f}{\partial c_i} + J = 0. \quad (2)$$

Here, $(-e)$ and $m_e$ are the electron charge (negative) and mass, respectively; $E_i$ and $B_i$ are the electric and magnetic field strengts, respectively; $c_i$ is the electron velocity; $\varepsilon_{ikl}$ is a fully antisymmetric Levi-Civita symbol; $c$ is the speed of light; and $F_i$ is the acceleration induced by an external force.

Let us introduce the electron velocity related to thermal motion, $v_i = c_i - c_{0i}$, where $c_{0i}$ is the average mass velocity. Note that the velocity $v_i$ is here equivalent to the chaotic velocity. The equilibrium distribution of nonrelativistic electrons in the comoving reference frame is described by the Fermi–Dirac function

$$f_0 = \left[1 + \exp\left(\frac{m_e v^2 - 2\mu}{2kT}\right)\right]^{-1},$$
$$R\int f_0 dv_i = n_e, \quad R = \frac{2m_e^3}{h^3}. \quad (3)$$

Here, $\mu$ is the chemical potential of electrons, $k$ is the Boltzmann constant, and $T$ is the temperature. The zero-order distribution function of nuclei $f_{N0}$ is assumed to be isotropic in velocities and dependent on the local thermodynamic parameters; otherwise, it can be arbitrary and normalized so that

$$n_N = \int f_{N0} dc_{Ni}, \quad (4)$$

where $n_N$ is the number density of the nuclei, $n_e = Zn_N$, and $Z$ is the nucleus charge.

In a gravitational field of a mass $M$, in spherical coordinates $(r, \phi, \theta)$, we have

$$F_{Gi} = \left(-\frac{GM}{r^2}, 0, 0\right). \quad (5)$$

According to [4, 15–17], the collision integral $J$ for an arbitrary degeneracy of electrons is given by

$$J = J_{ee} + J_{eN} = R \int \left[ f' f_1'(1-f)(1-f_1) - f f_1 (1-f')(1-f_1') \right] g_{ee} b \, db \, d\varepsilon \, dc_{1i} + \int \left[ f' f_N'(1-f) - f f_N (1-f') \right] \times g_{eN} b \, db \, d\varepsilon \, dc_{Ni}. \quad (6)$$

Here, $f$ and $f_1$ are the electron distribution functions depending on velocities $c_i$ and $c_{1i}$, respectively [4], while the impact parameters $b$ and $\varepsilon$ represent the geometric parameters of a collision of particles moving with relative velocities $g_{ee}$ and $g_{eN}$.

Integration in the electron part of the collision integral in expression (6) is performed over the velocity space of colliding particles ($dc_{1i}$) and their physical space ($b \, db \, d\varepsilon$) [4]. The functions corresponding to velocities after collision are marked with primes.

The Boltzmann equation for electrons with pair collision integral (6) can be used when the electron gas is considered to be nearly ideal, i.e., the kinetic energy of electrons is much higher that the energy of electrostatic interaction. This condition is satisfied in plasma of sufficiently low density. In neutron stars and white dwarfs, the conditions are opposite: the plasma is so dense that the electron degeneracy must be taken into account. It is known from statistical physics [18] that a gas of strongly degenerate electrons becomes ideal because, in this case, thermal energy is replaced by a large Fermi energy. Therefore, the results of calculations performed in the present work are valid for low-density plasma, as well as for dense plasma with degenerate electrons. Detailed discussion of the applicability of pair-collision integral (6) and its modifications for dense nondegenerate gases can be found in [4].

A collision integral for degenerate neutrons in nuclear matter, similar to $J_{ee}$ in expression (6), can be found in [15] (see also [19]). In the presence of nondegenerate heavy nuclei and strongly degenerate neutrons, the contribution of collisions between the latter to the thermal conductivity and diffusion coefficient is negligibly small compared to the contribution of neutron–nucleus collisions. Such a situation takes place for strongly degenerate electrons. Therefore, the Lorentz approximation with allowance for collisions between light and heavy particles is asymptotically accurate. Hence, in our case, we can neglect $J_{ee}$ compared to $J_{eN}$ and assume that $J = J_{eN}$ in expression (6).

The Boltzmann equation with respect to thermal velocity $v_i$ has the form [9]

$$\frac{df}{dt} + v_i \frac{\partial f}{\partial r_i} - \left[ \frac{e}{m_e}\left( E_i + \frac{1}{c}\varepsilon_{ikl} v_k B_l \right) - F_i + \frac{dc_{0i}}{dt} \right] \frac{\partial f}{\partial v_i} - \frac{e}{m_e c}\varepsilon_{ikl} v_k B_l \frac{\partial f}{\partial v_i} - \frac{\partial f}{\partial v_i} v_k \frac{\partial c_{0i}}{\partial r_k} + J = 0, \quad (7)$$

where

$$\frac{d}{dt} = \frac{\partial}{\partial t} + c_{0i} \frac{\partial}{\partial r_i}. \quad (8)$$

The transport equations for the density, total momentum, and energy of electrons in a two-component mixture of electrons and nuclei, derived in a standard way from the Boltzmann equation for a quasineutral plasma [4, 7–9], has the form

$$\frac{dn_e}{dt} + n_e \frac{\partial c_{0i}}{\partial r_i} + \frac{\partial}{\partial r_i}(n_e \langle v_i \rangle) = 0, \quad (9)$$

$$\rho \frac{dc_{0i}}{dt} = \rho F_i + \frac{1}{c}\varepsilon_{ikl} j_k B_l - \frac{\partial \Pi_{ik}}{\partial r_k}, \quad (10)$$

$$\frac{3}{2} k n_e \frac{dT}{dt} - \frac{3}{2} kT \frac{\partial}{\partial r_i}(n_e \langle v_i \rangle) + \frac{\partial q_{ei}}{\partial r_i} + \Pi^e_{ik} \frac{\partial c_{0i}}{\partial r_k} = j_i \left( E_i + \frac{1}{c}\varepsilon_{ikl} c_{0k} B_l \right) - \rho_e \langle v_i \rangle \left( \frac{dc_{0i}}{dt} - F_i \right). \quad (11)$$

Henceforth, we will neglect electron inertia and, accordingly, the last term in Eq. (11), which has the order of smallness of $\sim(m_e/m_i)$ in comparison to other terms. Here,

$$\Pi_{ik} = \sum_\alpha n_\alpha m_\alpha \langle v_i^\alpha v_k^\alpha \rangle, \quad \Pi^e_{ik} = n_\alpha m_\alpha \langle v_i v_k \rangle, \quad (12)$$

$$\langle v_{\alpha i} \rangle = \frac{R}{n_\alpha} \int f v_{\alpha i} dc_{\alpha i},$$
$$n_e = R \int f dc_{ei}, \quad \rho_e = m_e n_e, \quad (13)$$

$$c_{0i} = \frac{1}{\rho} \sum_\alpha \rho_\alpha \langle c_{ai} \rangle, \quad j_i = -n_e e \langle v_i \rangle, \quad (14)$$

$$q_i = \frac{1}{2} n_e m_e \langle v_e^2 v_{ei} \rangle = \frac{1}{2} m_e R \int f v_e^2 v_{ei} dc_{ei}, \quad (15)$$

$\langle v_{ei} \rangle$ is the average electron velocity in the comoving reference frame, $q_i$ is the electron heat flux, $j_i$ is the electron electric current, and summation is performed over electrons and nuclei. When the electron viscosity is neglected, we have $\Pi^e_{ik} = P_e \delta_{ik}$, where $P_e = n_e m_e \langle v^2 \rangle/3$ is the electron pressure. Here and below, we assume the average mass velocity to be equal to the average ion velocity, $c_{0i} = \langle c_{Ni} \rangle$. We also take into account the electric current and heat flux produced only by electrons and write $v_i \equiv v_{ei}$. The electron density $n_e$ in quasineutral plasma is uniquely related to the mass density $\rho$, which determined by the nuclei, having the atomic mass $A$ ($m_N = Am_p$) and charge number $Z$,

$$\rho = m_N n_N, \quad n_e = \frac{Z\rho}{m_N}. \qquad (16)$$

## 3. DERIVATION OF EQUATIONS FOR THE ELECTRON DISTRIBUTION FUNCTIONS IN THE FIRST APPROXIMATION

The Boltzmann equation can be solved by the Chapman–Enskog successive iteration method [4]. This method is used when the distribution functions are close to those in thermodynamic equilibrium, while the deviations from equilibrium are considered in the linear approximation.

In the zero-order approximation, the electron distribution in thermodynamic equilibrium is determined by the Fermi–Dirac function, which causes the collision integrals $J_{ee}$ and $J_{eN}$ in expression (6) to vanish.

Substitution of formula (3) into Eqs. (9)–(15) yields the following transport equations in the zero-order approximation, in which $\langle v_i \rangle = 0$, $q_i = 0$, and $\Pi_{ik} = (P_e + P_N)\delta_{ik}$:

$$n_e = 2\left(\frac{2kTm_e}{h^2}\right)^{3/2} G_{3/2}(x_0),$$

$$P_e = 2kT\left(\frac{2kTm_e}{h^2}\right)^{3/2} G_{5/2}(x_0), \qquad (17)$$

$$G_n(x_0) = \frac{1}{\Gamma(n)} \int_0^\infty \frac{x^{n-1}dy}{1+\exp(x-x_0)}, \quad x_0 = \frac{\mu}{kT}, \qquad (18)$$

where $G_n(x_0)$ are the Fermi integrals. Henceforth, we will use notation $G_n$ instead of $G_n(x_0)$, because the argument is always the same. In the first-order approximation, we seek the distribution function $f$ in the form

$$f = f_0[1 + \chi(1-f_0)]. \qquad (19)$$

An isotropic distribution function $f_{N0}$ leads to the relation

$$\frac{1}{n_N}\int v_{Ni}v_{Nk}f_{N0}dc_{Ni} = \delta_{ik}\frac{kT}{m_N}. \qquad (20)$$

The function $\chi$ allows representing the solution in the form [14]

$$\chi = -A_i\frac{\partial \ln T}{\partial r_i} - n_e D_i d_i \frac{G_{5/2}}{G_{3/2}}, \qquad (21)$$

$$d_i = \frac{\rho_N}{\rho}\frac{\partial \ln P_e}{\partial r_i} - \frac{P_e}{P_e\rho}\frac{1}{\partial r_i}\frac{\partial P_N}{\partial r_i} \\ + \frac{en_e}{P_e}\left(E_i + \frac{1}{c}\varepsilon_{ikl}c_{0k}B_l\right) - \frac{m_e}{kT}F_i. \qquad (22)$$

Plasma is assumed to be quasineutral. The functions $A_i$ and $D_i$ describe the heat flux and particle diffusion, respectively. Gravitational force (5) is irrelevant for electron diffusion; however, it can be critically important for diffusion of nuclei and nucleons (see, e.g., [20]). To take the gravitational force into account in diffusion equation (9), expression (5) should be written in Cartesian coordinates. Substitution of expression (21) into the equation for $\chi$ yields equations for $A_i$ and $D_i$ [4, 14]. It was demonstrated in [8, 9] that polar vectors $A_i$ and $D_i$ in the presence of a magnetic field with an axial vector $B_i$ can be sought for in the form

$$A_i = A^{(1)}v_i + A^{(2)}\varepsilon_{ijk}v_jB_k + A^{(3)}B_i(v_jB_j), \qquad (23)$$

$$D_i = D^{(1)}v_i + D^{(2)}\varepsilon_{ijk}v_jB_k + D^{(3)}B_i(v_jB_j), \qquad (24)$$

where $v_i$, $\varepsilon_{ijk}v_jB_k$, and $B_i(v_jB_j)$ are three linearly independent polar vectors and $A^{(\alpha)}$ and $D^{(\alpha)}$ ($\alpha = 1, 2, 3$) are functions of the scalars $v^2$ and $B^2$.

Introducing functions

$$\xi_A = A^{(1)} + iBA^{(2)}, \quad \xi_D = D^{(1)} + iBD^{(2)} \qquad (25)$$

and dimensionless velocity $u_i = v_i\sqrt{m_e/2kT}$, omitting small (compared to unity) terms on the order of $\sim m_e/m_N$, we obtain the equations for $\xi_A$ and $\xi_D$,

$$f_0(1-f_0)\left(u^2 - \frac{5G_{5/2}}{2G_{3/2}}\right)u_i \\ = -iBf_0(1-f_0)\frac{e\xi_A}{m_e c}u_i + I_{eN}(\xi_A u_{Ni}), \qquad (26)$$

$$f_0(1-f_0)u_i = -iBf_0(1-f_0)\frac{e\xi_D}{m_e c}u_i + I_{eN}(\xi_D u_{Ni}), \qquad (27)$$

where

$$I_{eN}(\xi u_{Ni}) \\ = \int f_0 f_{N0}(1-f_0')(\xi u_i - \xi' u_i')g_{eN}bdbd\varepsilon dc_{Ni}. \qquad (28)$$

## 4. THERMAL DIFFUSION, DIFFUSION, AND DIFFUSION THERMOEFFECT OF DEGENERATE ELECTRONS IN THE PRESENCE OF A MAGNETIC FIELD

General expressions for the heat flux $q_i$ and average directional (diffusion) electron velocity $\langle v_i \rangle$ are given by

$$q_i = -\lambda_{ij}\frac{\partial T}{\partial x_j} - n_e\frac{G_{5/2}}{G_{3/2}}\nu_{ij}d_j = q_i^{(A)} + q_i^{(D)}, \qquad (29)$$

$$\langle v_i \rangle = -\mu_{ij}\frac{\partial T}{\partial x_j} - n_e\frac{G_{5/2}}{G_{3/2}}\eta_{ij}d_j = \langle v_i^{(A)} \rangle + \langle v_i^{(D)} \rangle, \qquad (30)$$

where $\lambda_{ij}$ and $\nu_{ij}$ are the thermal conductivity and diffusion thermoeffect tensors, respectively, while $\mu_{ij}$ and $\eta_{ij}$ are the thermal diffusion and diffusion tensors, respectively [19, 21]. The indices $(A)$ and $(D)$ correspond to the heat flux and diffusion velocity of electrons determined by the temperature gradient $\partial T/\partial x_j$ and diffusion vector $d_j$, respectively.

The procedure of finding the components of the thermal conductivity tensor $\lambda_{ij}$ was described in detail in [14], where analytic expressions for them were derived. Similarly, to find the components of the tensors $\mu_{ij}$, $\nu_{ij}$, and $\eta_{ij}$, it is necessary to write Eqs. (26)–(28) for the functions $\xi_A$ and $\xi_D$ by using formulas (23)–(25) and relations $f_0' = f_0$, $\xi' = \xi$, $u_i' = u_i \cos\theta$ and integrating with respect to $dc_{Ni}$ with allowance for expression (4). Dividing Eqs. (26) and (27) by $f_0(1 - f_0)$, we can be recast them in the form

$$u^2 - \frac{5 G_{5/2}}{2 G_{3/2}}$$
$$= -i\frac{eB}{m_e c}\xi_A + n_N \xi_A \int (1 - \cos\theta) g_{eN} b\, db\, d\varepsilon, \quad (31)$$

$$\frac{1}{n_e} = -i\frac{eB}{m_e c}\xi_D + n_N \xi_D \int (1 - \cos\theta) g_{eN} b\, db\, d\varepsilon, \quad (32)$$

where Eq. (31) is used to calculate the components of the thermal conductivity tensor $\lambda_{ij}$ and diffusion thermoeffect tensor $\nu_{ij}$, while Eq. (32) is used to calculate the components of the diffusion tensor $\eta_{ij}$ and thermal diffusion tensor $\mu_{ij}$.

### 4.1. Thermal Diffusion of Degenerate Electrons

Let us find the components of the thermal diffusion tensor. Equation (31) yields the solution for the function $\xi_A$ in the form [14]

$$\xi_A = \frac{u^2 - \dfrac{5 G_{5/2}}{2 G_{3/2}}}{2\pi n_N \int_0^\infty (1 - \cos\theta) g_{eN} b\, db - i\omega}. \quad (33)$$

In the Lorentz approximation, at $g_{eN} = v$, integral in expression (33) can be calculated analytically [14],

$$\int_0^\infty (1 - \cos\theta) g_{eN} b\, db = 2\frac{e^4 Z^2}{m_e^2 v^3}\Lambda. \quad (34)$$

Here, $\Lambda$ is the Coulomb integral written in the form [22]

$$\Lambda = \ln\left(\frac{b_{\max}\overline{v}_e^2 m_e}{Ze^2}\right), \quad \Lambda \gg 1, \quad (35)$$

where

$$\overline{v}_e^2 = \frac{3kT}{m_e}\frac{G_{5/2}}{G_{3/2}} = \begin{cases} \dfrac{3kT}{m_e} & \text{(ND)} \\ \dfrac{3}{5}\dfrac{h^2}{m_e^2}\left(\dfrac{3n_e}{8\pi}\right)^{2/3} & \text{(D)}. \end{cases} \quad (36)$$

Notations "ND" and "D" correspond to nondegenerate and strongly degenerate electrons, respectively. The quantity $b_{\max}$ represents the Debye length, which can be expressed via the electron and ion the Debye lengths ($r_{De}$ and $r_{Di}$) as

$$\frac{1}{b_{\max}^2} = \frac{1}{r_{Di}^2} + \frac{1}{r_{De}^2} = \frac{4\pi e^2}{kT}\left(n_N Z^2 + n_e\frac{G_{1/2}}{G_{3/2}}\right), \quad (37)$$

where

$$\frac{G_{1/2}}{G_{3/2}} = \begin{cases} 1 & \text{(ND)} \\ 4(3\pi^2)^{1/3}\dfrac{m_e kT}{h^2 n_e^{2/3}} & \text{(D)} \end{cases}. \quad (38)$$

The influence of quantum effects on Debye screening was discussed in [3].

The average frequency of electron–ion collisions $\nu_{ei}$ was expressed in [23] in the form

$$\nu_{ei} = \frac{4}{3}\sqrt{\frac{2\pi}{m_e}}\frac{Z^2 e^4 n_N \Lambda}{(kT)^{3/2} G_{3/2}}\frac{1}{1 + e^{-x_0}}, \quad \tau_{ei} = 1/\nu_{ei}. \quad (39)$$

In the limiting cases, it can be written as

$$\nu_{ei} = \begin{cases} \dfrac{4}{3}\sqrt{\dfrac{2\pi}{m_e}}\dfrac{Z^2 e^4 n_N \Lambda}{(kT)^{3/2}} & \text{(ND)}, \quad \tau_{nd} = 1/\nu_{nd}, \\ \dfrac{32\pi^2}{3} m_e \dfrac{Z^2 e^4 \Lambda n_N}{h^3 n_e} & \text{(D)}, \quad \tau_d = 1/\nu_d. \end{cases} \quad (40)$$

The functions $\xi_A$, $A^{(1)}$, and $A^{(2)}$ in formulas (25) are given by [14]

$$\xi_A = \left(u^2 - \frac{5 G_{5/2}}{2 G_{3/2}}\right)\left[4\pi n_N \left(\frac{m_e}{2kT}\right)^{3/2}\frac{e^4 Z^2}{m_e^2 u^3}\Lambda - i\omega\right]^{-1}, \quad (41)$$

$$A^{(1)} = \left(u^2 - \frac{5 G_{5/2}}{2 G_{3/2}}\right)4\pi n_N \left(\frac{m_e}{2kT}\right)^{3/2}$$
$$\times \frac{e^4 Z^2}{m_e^2 u^3}\Lambda \left[\left(4\pi n_N \left(\frac{m_e}{2kT}\right)^{3/2}\frac{e^4 Z^2}{m_e^2 u^3}\Lambda\right)^2 + \omega^2\right]^{-1}, \quad (42)$$

$$A^{(2)} = \frac{\omega}{B}\left(u^2 - \frac{5 G_{5/2}}{2 G_{3/2}}\right)$$
$$\times \left[\left(4\pi n_N \left(\frac{m_e}{2kT}\right)^{3/2}\frac{e^4 Z^2}{m_e^2 u^3}\Lambda\right)^2 + \omega^2\right]^{-1}. \quad (43)$$

The function $A^{(3)}$ in formula (23) can be expressed in the form

$$B^2 A^{(3)} = A^{(1)}(B=0) - A^{(1)}. \quad (44)$$

The expression for the average velocity related to thermal diffusion that follows from Eqs. (13), (19), (21), (25), (30), and (41)–(44) has the form

$$\langle v_i^{(A)} \rangle = -\left(\mu^{(1)}\delta_{ij} + \mu^{(2)}\varepsilon_{ijk}B_k + \mu^{(3)}B_iB_j\right)\frac{\partial T}{\partial x_j}, \quad (45)$$

where

$$\mu^{(1)} = \frac{4\pi}{3}\frac{m_e^3}{h^3 T n_e}\left(\frac{2kT}{m_e}\right)^{5/2}\int_0^\infty f_0(1-f_0)A^{(1)}y^{3/2}dy, \quad (46)$$

$$\mu^{(2)} = -\frac{4\pi}{3}\frac{m_e^3}{h^3 T n_e}\left(\frac{2kT}{m_e}\right)^{5/2}\int_0^\infty f_0(1-f_0)A^{(2)}y^{3/2}dy, \quad (47)$$

$$\mu^{(3)} = \frac{4\pi}{3}\frac{m_e^3}{h^3 T n_e}\left(\frac{2kT}{m_e}\right)^{5/2}\int_0^\infty f_0(1-f_0)A^{(3)}y^{3/2}dy, \quad (48)$$

$$y = u^2.$$

For strongly degenerate electrons at $x_0 \gg 1$, integrals in Eqs. (46)–(48), where expressions for $A^{(1)}$, $A^{(2)}$, and $A^{(3)}$ are given by formulas (42)–(44), can be expressed analytically by using expansions [14, 18]

$$\int_0^\infty \frac{f(x)dx}{e^{x-x_0}+1} = \int_0^\infty f(x)dx + \frac{\pi^2}{6}f'(x_0) + \ldots,$$

$$G_n(x_0)\frac{1}{\Gamma(n)}\left[\frac{x^n}{n} + \frac{\pi^2}{6}(n-1)x_0^{n-2} + \ldots\right], \quad (49)$$

$$x_0 = \frac{\mu}{kT} \cong \frac{\varepsilon_{fe}}{kT} = \frac{(3\pi^2 n_e)^{2/3}h^2}{8\pi^2 m_e kT} \gg 1.$$

After partial integration, we obtain the following expressions suitable for integration according to formulas (49):

$$\mu^{(1)} = \frac{4\pi}{3}\frac{m_e^3}{h^3 T n_e}\left(\frac{2kT}{m}\right)^{5/2}\int_0^\infty f_0\frac{d(A^{(1)}y^{3/2})}{dy}dy, \quad (50)$$

$$\mu^{(2)} = -\frac{4\pi}{3}\frac{m_e^3}{h^3 T n_e}\left(\frac{2kT}{m}\right)^{5/2}\int_0^\infty f_0\frac{d(A^{(2)}y^{3/2})}{dy}dy, \quad (51)$$

$$B^2 A^{(3)} = A^{(1)}(B=0) - A^{(1)}. \quad (52)$$

Applying expansions (49) to integrals (50) and (51), we have

$$\mu^{(1)} = \frac{4\pi}{3}\frac{m_e^3}{h^3 T n_e}\left(\frac{2kT}{m_e}\right)^{5/2}$$
$$\times\left[A^{(1)}(x_0)x_0^{3/2} + \frac{\pi^2}{6}\frac{d^2(A^{(1)}y^{3/2})}{dy^2}\bigg|_{y=x_0}\right], \quad (53)$$

$$\mu^{(2)} = -\frac{4\pi}{3}\frac{m_e^3}{h^3 T n_e}\left(\frac{2kT}{m_e}\right)^{5/2}$$
$$\times\left[A^{(2)}(x_0)x_0^{3/2} + \frac{\pi^2}{6}\frac{d^2(A^{(2)}y^{3/2})}{dy^2}\bigg|_{y=x_0}\right], \quad (54)$$

$$B^2\mu^{(3)} = \mu^{(1)}(B=0) - \mu^{(1)}. \quad (55)$$

The average time between electron–ion collisions $\tau_{ei}$ is equal to the reciprocal of the electron–ion collision frequency $\nu_{ei}$ (see Eqs. (39), (40)). Using formulas (42) and (43) and writing expressions in which $\tau_d$ is given by the reciprocal of $\nu_{ei}$ for the degenerate case, let us write out the functions $A^{(1)}$ and $A^{(2)}$ and the components of the thermal diffusion tensor in the form

$$A^{(1)} = \tau_d \frac{y^{3/2}}{x_0^{3/2}} \frac{y - \frac{5}{2}\frac{G_{5/2}}{G_{3/2}}}{1+\omega^2\tau_d^2(y^3/x_0^3)},$$

$$A^{(2)} = \frac{\omega\tau_d^2}{B}\frac{y^3}{x_0^3}\frac{y - \frac{5}{2}\frac{G_{5/2}}{G_{3/2}}}{1+\omega^2\tau_d^2(y^3/x_0^3)}, \quad (56)$$

$$\mu^{(1)} = \frac{4\pi^3}{3}\frac{k^2 T}{n_e h^2}\left(\frac{3n_e}{\pi}\right)^{1/3}\tau_d\left[\frac{1}{1+\omega^2\tau_d^2}\right.$$
$$\left. - 2\frac{\omega^2\tau_d^2}{(1+\omega^2\tau_d^2)^2} - \frac{\pi^2}{6}\left(\frac{1}{1+\omega^2\tau_d^2(y^3/x_0^3)}\right)''\bigg|_{y=x_0}\right], \quad (57)$$

$$\mu^{(2)} = -\frac{4\pi^3}{3}\frac{k^2 T}{n_e h^2}\left(\frac{3n_e}{\pi}\right)^{1/3}\frac{\omega\tau_d^2}{B}\left[\frac{2}{1+\omega^2\tau_d^2}\right.$$
$$\left. - 2\frac{\omega^2\tau_d^2}{(1+\omega^2\tau_d^2)^2} - \frac{\pi^2}{6}\left(\frac{1}{1+\omega^2\tau_d^2(y^3/x_0^3)}\right)''\bigg|_{y=x_0}\right], \quad (58)$$

$$B^2\mu^{(3)} = \mu^{(1)}(B=0) - \mu^{(1)}. \quad (59)$$

Double prime in the last terms in expressions (57) and (58) denotes the second derivative with respect to $y$. After differentiation, they turn out to be of higher order of smallness; therefore, their contribution was neglected in the dependence plotted in Fig. 1. Henceforth, it will become evident that taking into account small terms $\sim 1/x_0^2$ increases the difference between the rigorous expressions and approximation (1).

### 4.2. Diffusion of Degenerate Electrons

To calculate the components of the diffusion and diffusion thermoeffect tensors from Eq. (32), let us write out an expression for $\xi_D$,

$$\xi_D = \frac{1}{n_e} \frac{1}{2\pi n_N \int (1 - \cos\theta) g b db - i\omega}. \tag{60}$$

Taking into account Eqs. (34) and (40), after integration, we obtain the following expressions for the functions $\xi_D$, $D^{(1)}$, and $D^{(2)}$ in formulas (25):

$$\xi_D = \frac{1}{n_e}\left[4\pi n_N \frac{e^4 Z^2}{m^2 u^3}\Lambda\left(\frac{m_e}{2kT}\right)^{3/2} - i\omega\right]^{-1}, \tag{61}$$

$$D^{(1)} = \frac{1}{n_e} 4\pi n_N \frac{e^4 Z^2}{m^2 u^3}\Lambda\left(\frac{m_e}{2kT}\right)^{3/2}$$
$$\times\left[\left(4\pi n_N \frac{e^4 Z^2}{m^2 u^3}\Lambda\left(\frac{m_e}{2kT}\right)^{3/2}\right)^2 + \omega^2\right]^{-1} \tag{62}$$
$$= \frac{\tau_d}{n_e}\frac{y^{3/2}}{x_0^{3/2}}\frac{1}{1+\omega^2\tau_d^2(y^3/x_0^3)},$$

$$D^{(2)} = \frac{1}{n_e}\frac{\omega}{B}\left[\left(4\pi n_N \frac{e^4 Z^2}{m^2 u^3}\Lambda\left(\frac{m_e}{2kT}\right)^{3/2}\right)^2 + \omega^2\right]^{-1}$$
$$= \frac{\omega\tau_d^2}{n_e B}\frac{y^3}{x_0^3}\frac{1}{1+\omega^2\tau_d^2(y^3/x_0^3)}, \tag{63}$$

$$B^2 D^{(3)} = D^{(1)}\big|_{B=0} - D^{(1)}. \tag{64}$$

The average electron velocity $\langle v_i^{(D)}\rangle$ related to diffusion is determined by the tensor $\eta_{ij}$,

$$\langle v_i^{(D)}\rangle = -n_e \frac{G_{5/2}}{G_{3/2}}\left(\eta^{(1)}\delta_{ij} + \eta^{(2)}\varepsilon_{ijk}B_k + \eta^{(3)}B_i B_j\right)d_j, \tag{65}$$

where

$$\eta^{(1)} = \frac{4\pi}{3}\frac{m_e^3}{h^3 n_e}\left(\frac{2kT}{m_e}\right)^{5/2}\int_0^\infty f_0(1-f_0)D^{(1)}y^{3/2}dy, \tag{66}$$

$$\eta^{(2)} = -\frac{4\pi}{3}\frac{m_e^3}{h^3 n_e}\left(\frac{2kT}{m_e}\right)^{5/2}\int_0^\infty f_0(1-f_0)D^{(2)}y^{3/2}dy, \tag{67}$$

$$\eta^{(3)} = \frac{4\pi}{3}\frac{m_e^3}{h^3 n_e}\left(\frac{2kT}{m_e}\right)^{5/2}\int_0^\infty f_0(1-f_0)D^{(3)}y^{3/2}dy. \tag{68}$$

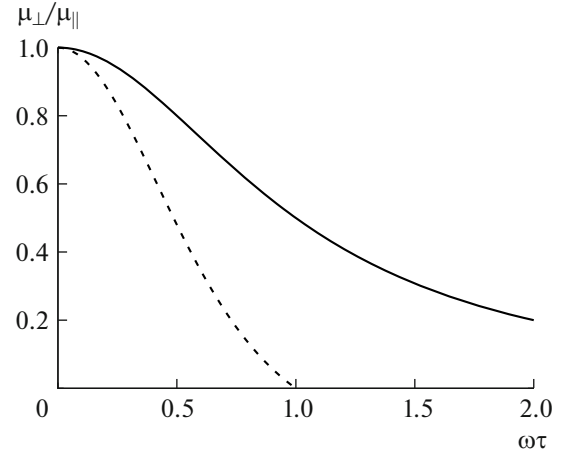

**Fig. 1.** Ratio $\mu_\perp/\mu_\parallel$ as a function of $\omega\tau$. For comparison, the curves representing the phenomenologically determined thermal diffusion coefficient (solid curve) and that obtained from the asymptotic solution to the Boltzmann equation (dashed curve) are shown.

Using expansions (49), the components of the diffusion tensor can be recast in the form

$$\eta^{(1)} = \frac{4\pi}{3}\frac{m_e^3}{h^3 n_e}\left(\frac{2kT}{m_e}\right)^{5/2}$$
$$\times\left(D^{(1)}(x_0)x_0^{3/2} + \frac{\pi^2}{6}\frac{d^2 D^{(1)}(y)y^{3/2}}{dy^2}\bigg|_{y=x_0}\right), \tag{69}$$

$$\eta^{(2)} = -\frac{4\pi}{3}\frac{m_e^3}{h^3 n_e}\left(\frac{2kT}{m_e}\right)^{5/2}$$
$$\times\left(D^{(2)}(x_0)x_0^{3/2} + \frac{\pi^2}{6}\frac{d^2 D^{(2)}(y)y^{3/2}}{dy^2}\bigg|_{y=x_0}\right), \tag{70}$$

$$B^2\eta^{(3)} = \eta^{(1)}\big|_{B=0} - \eta^{(1)}. \tag{71}$$

Taking into account the expression for $\tau_d$ in formula (40), along with formulas (62) and (63), we can write expressions for the components of the diffusion tensor in the form

$$\eta^{(1)} = \frac{kT}{n_e m_e}$$
$$\times \tau_d\left(\frac{1}{1+\omega^2\tau_d^2} + \frac{\pi^2}{6}\left(\frac{1}{1+\omega^2\tau_d^2(y^3/x_0^3)}\right)''\bigg|_{x=x_0}\right), \tag{72}$$

$$\eta^{(2)} = -\frac{kT}{n_e m_e}\frac{\omega\tau_d^2}{B}$$
$$\times\left(\frac{1}{1+\omega^2\tau_d^2} + \frac{\pi^2}{6}\left(\frac{1}{1+\omega^2\tau_d^2(y^3/x_0^3)}\right)''\bigg|_{x=x_0}\right), \tag{73}$$

$$B^2\eta^{(3)} = \eta^{(1)}(B=0) - \eta^{(1)}. \tag{74}$$

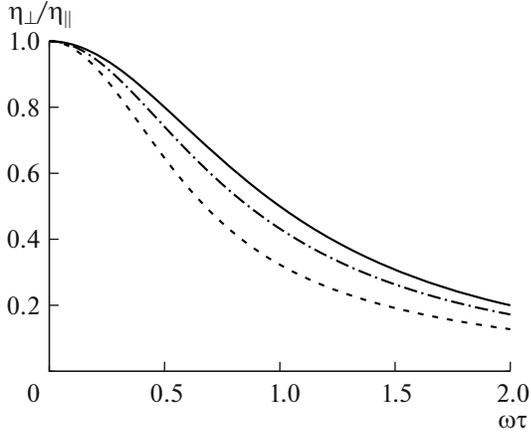

**Fig. 2.** Ratio $\eta_\perp/\eta_\parallel$ as a function of $\omega\tau$. For diffusion, the phenomenologically obtained solid curve coincides with the curve obtained by solving the Boltzmann equation in the case of strong degeneracy. If terms on the order of smallness of $1/x_0^2$ are retained in the exact solution, then the solution obtained from the Boltzmann equation differs from the phenomenological one given by Eq. (1). The dash-dotted and dashed curves correspond to the solutions obtained with allowance for small terms at $kT = 0.11 E_f$ ($x_0 = 9$) and $kT = 0.2 E_f$ ($x_0 = 5$), respectively.

After differentiation, the last terms in expressions (72) and (73) turn out to be of higher order of smallness; therefore, their contribution was neglected when plotting the dependence in Fig. 2. In this case, the exact asymptotic solution coincides with the approximate one given by Eq. (1). Taking into account small terms $\sim 1/x_0^2$ leads to a deviation of the kinetic solution from approximation (1) in Fig. 2.

### 4.3. Diffusion Thermoeffect of Degenerate Electrons

The components of the diffusion thermoeffect tensor can be calculated using relations (62) and (63), by analogy with the calculation of the components of the diffusion tensor. Heat flux $q_i^{(D)}$ due to the diffusion thermoeffect can be expressed as

$$q_i^{(D)} = -n_e \frac{G_{5/2}}{G_{3/2}} \left( \nu^{(1)} \delta_{ij} + \nu^{(2)} \varepsilon_{ijk} B_k + \nu^{(3)} B_i B_j \right) d_j, \quad (75)$$

where

$$\nu^{(1)} = \frac{2\pi}{3} \frac{m_e^4}{h^3} \left( \frac{2kT}{m_e} \right)^{7/2} \int_0^\infty f_0(1-f_0) D^{(1)} y^{5/2} dy, \quad (76)$$

$$\nu^{(2)} = -\frac{2\pi}{3} \frac{m_e^4}{h^3} \left( \frac{2kT}{m_e} \right)^{7/2} \int_0^\infty f_0(1-f_0) D^{(2)} y^{5/2} dy, \quad (77)$$

$$\nu^{(3)} = \frac{2\pi}{3} \frac{m_e^4}{h^3} \left( \frac{2kT}{m_e} \right)^{7/2} \int_0^\infty f_0(1-f_0) D^{(3)} y^{5/2} dy. \quad (78)$$

Using expansions (49), the components of the diffusion thermoeffect tensor can be written in the form

$$\nu^{(1)} = \frac{2\pi}{3} \frac{m_e^4}{h^3} \left( \frac{2kT}{m_e} \right)^{7/2}$$
$$\times \left( D^{(1)}(x_0) x_0^{5/2} + \frac{\pi^2}{6} \frac{d^2 D^{(1)}(y) y^{5/2}}{dy^2} \bigg|_{y=x_0} \right), \quad (79)$$

$$\nu^{(2)} = -\frac{2\pi}{3} \frac{m_e^4}{h^3} \left( \frac{2kT}{m_e} \right)^{7/2}$$
$$\times \left( D^{(2)}(x_0) x_0^{5/2} + \frac{\pi^2}{6} \frac{d^2 D^{(2)}(y) y^{5/2}}{dy^2} \bigg|_{y=x_0} \right), \quad (80)$$

$$B^2 \nu^{(3)} = \nu^{(1)}(B=0) - \nu^{(1)}. \quad (81)$$

Integration yields the following expressions for the components of the diffusion thermoeffect tensor:

$$\nu^{(1)} = \frac{kTh^2}{8m_e^2} \left( \frac{3n_e}{\pi} \right)^{2/3}$$
$$\times \tau_d \left( \frac{1}{1+\omega^2 \tau_d^2} + \frac{\pi^2}{6} \left( \frac{1}{1+\omega^2 \tau_d^2 (y^3/x_0^3)} \right)'' \bigg|_{x=x_0} \right), \quad (82)$$

$$\nu^{(2)} = -\frac{kTh^2}{8m_e^2} \left( \frac{3n_e}{\pi} \right)^{2/3} \frac{\omega \tau_d^2}{B}$$
$$\times \left( \frac{1}{1+\omega^2 \tau_d^2} + \frac{\pi^2}{6} \left( \frac{1}{1+\omega^2 \tau_d^2 (y^3/x_0^3)} \right)'' \bigg|_{x=x_0} \right), \quad (83)$$

$$B^2 \nu^{(3)} = \nu^{(1)}(B=0) - \nu^{(1)}. \quad (84)$$

After differentiation, the last terms in expressions (82) and (83) turn out to be of higher order of smallness; therefore, their contribution was neglected when plotting the dependences in Fig. 3. In this case, the exact asymptotic solution coincides with the approximate one given by Eq. (1). Taking into account small terms $\sim 1/x_0^2$ leads to a deviation of the kinetic solution from approximation (1) in Fig. 3.

For the sake of completeness, we reproduce here expressions for the components of the tensor $\lambda_{ij}$ in a strongly degenerate plasma [14]:

$$\lambda^{(1)} = \frac{5\pi^2}{6} \frac{k^2 T n_e}{m_e} \tau_d \left\{ \frac{1}{1+\omega^2 \tau_d^2} - \frac{6}{5} \frac{\omega^2 \tau_d^2}{(1+\omega^2 \tau_d^2)^2} \right.$$
$$\left. - \frac{\pi^2}{10} \left[ \frac{1}{1+\omega^2 \tau_d^2 (y^3/x_0^3)} \right]'' \bigg|_{x=x_0} \right\}, \quad (85)$$

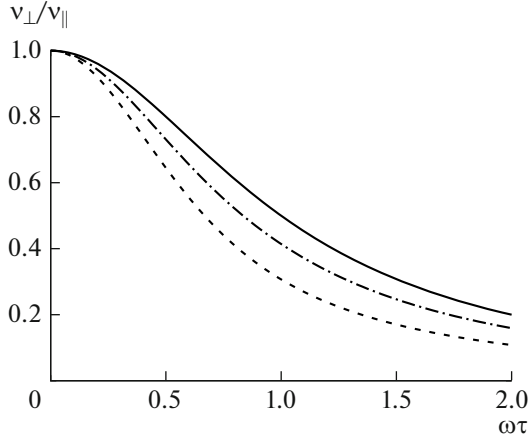

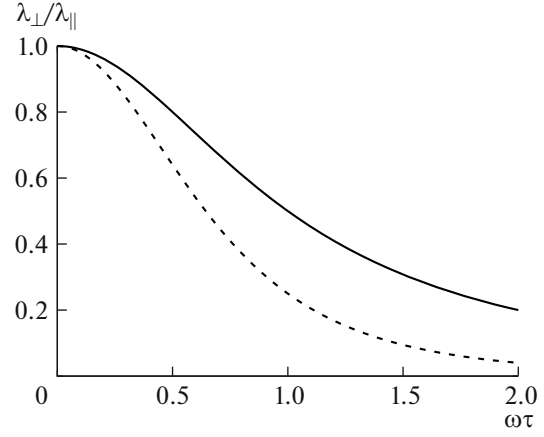

**Fig. 3.** Ratio $\nu_\perp/\nu_\parallel$ as a function of $\omega\tau$. Similar to diffusion, the phenomenologically obtained solid curve coincides with the asymptotic solution for the diffusion thermoeffect. If terms on the order of smallness of $1/x_0^2$ are retained in the solution obtained from the Boltzmann equation, then the plot of $\nu_\perp/\nu_\parallel$ differs from the phenomenological curve. The dash-dotted and dashed curves correspond to the solutions obtained at $kT = 0.11 E_f$ ($x_0 = 9$) and $kT = 0.2 E_f$ ($x_0 = 5$), respectively.

**Fig. 4.** Ratio $\lambda_\perp/\lambda_\parallel$ as a function of $\omega\tau$. For comparison, the curves representing the phenomenologically determined thermal conductivity (solid curve) and the asymptotic thermal conductivity obtained by solving the Boltzmann equation (dashed curve) are shown.

$$\lambda^{(2)} = -\frac{4\pi^2}{3}\frac{k^2 T n_e}{m_e}\frac{\tau_d^2 \omega}{B}\left\{\frac{1}{1+\omega^2\tau_d^2} - \frac{3}{4}\frac{\omega^2\tau_d^2}{(1+\omega^2\tau_d^2)^2}\right.$$

$$\left.-\frac{\pi^2}{16}\left[\frac{1}{1+\omega^2\tau_d^2(y^3/x_0^3)}\right]''\bigg|_{x=x_0}\right\}, \quad (86)$$

$$B^2 \lambda^{(3)} = \lambda^{(1)}(B=0) - \lambda^{(1)}. \quad (87)$$

The last terms in expressions (85) and (86) are small at $x_0^2 \gg 1$ and were not taken into consideration when plotting the dependences presented in Fig. 4.

Having calculated all four transport coefficients, taking into account Eqs. (29), (30), (45), (65), and (75), and using the thermal conductivity tensor from [14], we obtain the following expressions for the heat flux and the electron diffusion flux:

$$q_i = q_i^{(A)} + q_i^{(D)}$$
$$= -\left(\lambda^{(1)}\delta_{ij} + \lambda^{(2)}\varepsilon_{ijk}B_k + \lambda^{(3)}B_i B_j\right)\frac{\partial T}{\partial x_j} \quad (88)$$
$$- n_e \frac{G_{5/2}}{G_{3/2}}\left(\nu^{(1)}\delta_{ij} + \nu^{(2)}\varepsilon_{ijk}B_k + \nu^{(3)}B_i B_j\right)d_j,$$

$$\langle v_i \rangle = \langle v_i^{(A)} \rangle + \langle v_i^{(D)} \rangle$$
$$= -n_e \frac{G_{5/2}}{G_{3/2}}\left(\eta^{(1)}\delta_{ij} + \eta^{(2)}\varepsilon_{ijk}B_k + \eta^{(3)}B_i B_j\right)d_j \quad (89)$$
$$-\left(\mu^{(1)}\delta_{ij} + \mu^{(2)}\varepsilon_{ijk}B_k + \mu^{(3)}B_i B_j\right)\frac{\partial T}{\partial x_j}.$$

Numerical calculations require using equations written in the coordinate form. As an example, let us write out the components of the heat flux vector $q_i^{(A)}$ in Cartesian coordinates $x$, $y$, and $z$,

$$q_x^{(A)} = -\left[\lambda^{(1)}\frac{\partial T}{\partial x}\right.$$
$$+ \lambda^{(3)}B_x\left(B_x\frac{\partial T}{\partial x} + B_y\frac{\partial T}{\partial y} + B_z\frac{\partial T}{\partial z}\right) \quad (90)$$
$$\left.+ \lambda^{(2)}\left(B_z\frac{\partial T}{\partial y} - B_y\frac{\partial T}{\partial z}\right)\right],$$

$$q_y^{(A)} = -\left[\lambda^{(1)}\frac{\partial T}{\partial y}\right.$$
$$+ \lambda^{(3)}B_y\left(B_x\frac{\partial T}{\partial x} + B_y\frac{\partial T}{\partial y} + B_z\frac{\partial T}{\partial z}\right) \quad (91)$$
$$\left.+ \lambda^{(2)}\left(B_x\frac{\partial T}{\partial z} - B_z\frac{\partial T}{\partial x}\right)\right],$$

$$q_z^{(A)} = -\left[\lambda^{(1)}\frac{\partial T}{\partial z}\right.$$
$$+ \lambda^{(3)}B_z\left(B_x\frac{\partial T}{\partial x} + B_y\frac{\partial T}{\partial y} + B_z\frac{\partial T}{\partial z}\right) \quad (92)$$
$$\left.+ \lambda^{(2)}\left(B_y\frac{\partial T}{\partial x} - B_x\frac{\partial T}{\partial y}\right)\right].$$

The tensor structures of the expressions in brackets in Eqs. (88) and (89) are similar. Therefore, the components of other kinetic coefficients, $\nu^{(\alpha)}$, $\eta^{(\alpha)}$, and $\mu^{(\alpha)}$, where $\alpha = 1, 2, 3$, are described by expressions of the same form.

Set of MHD equations (9)–(11) should be complemented with an equation for the magnetic field, which follows from Maxwell's equations

$$\nabla \times \mathbf{E} = -\frac{1}{c}\frac{\partial \mathbf{B}}{\partial t}, \quad \nabla \cdot \mathbf{B} = 0, \quad \nabla \times \mathbf{B} = \frac{4\pi}{c}\mathbf{j}. \quad (93)$$

In the case of a scalar conductivity $\sigma$, when

$$\mathbf{j} = \sigma\left(\mathbf{E} + \frac{1}{c}\mathbf{v} \times \mathbf{B}\right), \quad (94)$$

the equation for the magnetic field has the form [24]

$$\frac{\partial \mathbf{B}}{\partial t} = \nabla \times (\mathbf{v} \times \mathbf{B}) + \frac{c^2}{4\pi\sigma}\Delta \mathbf{B}. \quad (95)$$

The equations for the magnetic field are substantially more complicated when the kinetic effects are taken into account rigorously. Instead of Eq. (94), the expression for vector $\mathbf{j}$ following from Eqs. (14), (30), and (89) has the form

$$j_i = -en_e \langle v_i \rangle$$
$$= \frac{2}{5}en_e^2 x_0 \left(\eta^{(1)}\delta_{ij} + \eta^{(2)}\varepsilon_{ijk}B_k + \eta^{(3)}B_iB_j\right)d_j \quad (96)$$
$$+ en_e \left(\mu^{(1)}\delta_{ij} + \mu^{(2)}\varepsilon_{ijk}B_k + \mu^{(3)}B_iB_j\right)\frac{\partial T}{\partial x_j}.$$

The tensor components $\eta^{(k)}$ are given by expressions (72)–(74), while the components $\mu^{(k)}$ are determined by expressions (57)–(59). Here, we took into account that, in the case of strong degeneracy, expansions (49) yield the relation $G_{5/2}/G_{3/2} \approx 2x_0/5$. Retaining the main terms, Eq. (22) for the vector $d_i$ can be transformed into the equation

$$d_i = \frac{\partial \ln P_e}{\partial r_i} + \frac{en_e}{P_e}\left(E_i + \frac{1}{c}\varepsilon_{ikl}c_{0k}B_l\right) - \frac{m_e}{kT}F_i. \quad (97)$$

In a general case, Eqs. (93), (96), and (97) should be used instead of Eqs. (93)–(95) to find the electric and magnetic fields. Equation (95) follows from expressions (93), (96), and (97) if we neglect thermal diffusion, take into account only generation of currents by the electromagnetic field, and neglect the dependence of the electric conductivity on the magnetic field, i.e., in the case of a scalar thermal conductivity given by

$$\sigma = \frac{2e^2n_e^3}{5P_e}x_0\eta, \quad \eta = \eta^{(1)}(B = 0).$$

The kinetic coefficients in a magnetic field that determine local heat and diffusion fluxes along ($q_\parallel$ and $j_\parallel$) and across ($q_\perp$ and $j_\perp$) the magnetic field, as well as the fluxes perpendicular to the plane defined by the magnetic field vector $B_i$ and any of the vectors $\partial T/\partial x_i$ or $d_i$, are frequently used in applications. These fluxes are referred to as the Hall fluxes, $q_{\text{Hall}}$ and $j_{\text{Hall}}$. To find expressions for the local fluxes, let us analyze the case in which the magnetic field is directed along the $x$ axis and there is a temperature gradient in the plane $(x, y)$, so that

$$B_i = (B_x, 0, 0), \quad \frac{\partial T}{\partial x_i} = \left(\frac{\partial T}{\partial x}, \frac{\partial T}{\partial y}, 0\right). \quad (98)$$

Equation (88) describing the heat flux yields

$$q_\parallel = q_x = -(\lambda^{(1)} + B^2\lambda^{(3)})\frac{\partial T}{\partial x} - (\nu^{(1)} + B^2\nu^{(3)})d_x,$$
$$\lambda_\parallel = \lambda^{(1)} + B^2\lambda^{(3)}, \quad \nu_\parallel = \nu^{(1)} + B^2\nu^{(3)}, \quad (99)$$

$$q_\perp = q_y = -\lambda^{(1)}\frac{\partial T}{\partial y} - \nu^{(1)}d_y,$$
$$\lambda_\perp = \lambda^{(1)}, \quad \nu_\perp = \nu^{(1)}, \quad (100)$$

$$q_{\text{Hall}} = q_z = \lambda^{(2)}B_x\frac{\partial T}{\partial y} + \nu^{(2)}B_xd_y,$$
$$\lambda_{\text{Hall}} = \lambda^{(2)}, \quad \nu_{\text{Hall}} = \nu^{(2)}. \quad (101)$$

It follows from Eq. (89) describing the average velocity of electrons that

$$\langle v_\parallel \rangle = \langle v_x \rangle$$
$$= -(\eta^{(1)} + B^2\eta^{(3)})d_x - (\mu^{(1)} + B^2\mu^{(3)})\frac{\partial T}{\partial x}, \quad (102)$$
$$\eta_\parallel = \eta^{(1)} + B^2\eta^{(3)}, \quad \mu_\parallel = \mu^{(1)} + B^2\mu^{(3)},$$

$$\langle v_\perp \rangle = \langle v_y \rangle = -\eta^{(1)}d_y - \mu^{(1)}\frac{\partial T}{\partial y},$$
$$\eta_\perp = \eta^{(1)}, \quad \mu_\perp = \mu^{(1)}, \quad (103)$$

$$\langle v_{\text{Hall}} \rangle = \langle v_z \rangle = \eta^{(2)}B_xd_y + \mu^{(2)}B_x\frac{\partial T}{\partial y},$$
$$\eta_{\text{Hall}} = \eta^{(2)}, \quad \mu_{\text{Hall}} = \mu^{(2)}. \quad (104)$$

Taking into account that $j_i = -n_ee\langle v_i \rangle$, the electric current of electrons in plasma can be expressed in the form

$$j_\parallel = en_e\left(\eta_\parallel d_x + \mu_\parallel \frac{\partial T}{\partial x}\right),$$
$$j_\perp = en_e\left(\eta_\perp d_y + \mu_\perp \frac{\partial T}{\partial y}\right), \quad (105)$$
$$j_{\text{Hall}} = -en_e\left(\eta_{\text{Hall}}B_xd_y + \mu_{\text{Hall}}B_x\frac{\partial T}{\partial y}\right).$$

In the case of strongly degenerate electrons, the above expressions for all coefficients of heat and electric charge transport are asymptotically exact, because electron–electron collisions can be neglected.

The quantity $[v^{(1)}]/[v^{(1)}(B=0)]$, which serves as an estimate of the influence of the magnetic field on the diffusion thermoeffect, is plotted in Fig. 3. Taking into account the terms on the order of smallness of $\sim 1/x_0^2$, expressions (72) and (73) describing diffusion can be written in a more detailed form as

$$\eta^{(1)} = \frac{kT}{n_e m_e}\tau_d\left(\frac{1}{1+\omega^2\tau_d^2} + \frac{1}{x_0^2}\frac{\pi^2}{1+\omega^2\tau_d^2}\right.$$
$$-\frac{1}{x_0^2}\frac{3\pi^2\omega^2\tau^2}{(1+\omega^2\tau_d^2)^2} \qquad (106)$$
$$\left.+\frac{\pi^2}{6}\left(\frac{1}{1+\omega^2\tau_d^2(y^3/x_0^3)}\right)''\bigg|_{x=x_0}\right),$$

$$\eta^{(2)} = -\frac{kT}{n_e m_e}\frac{\omega\tau_d^2}{B}\left(\frac{1}{1+\omega^2\tau_d^2}\right.$$
$$+\frac{21\pi^2}{8x_0^2}\frac{1}{1+\omega^2\tau^2} - \frac{9\pi^2}{2x_0^2}\frac{\omega^2\tau^2}{(1+\omega^2\tau^2)^2} \qquad (107)$$
$$\left.+\frac{\pi^2}{6}\left(\frac{1}{1+\omega^2\tau_d^2(y^3/x_0^3)}\right)''\bigg|_{x=x_0}\right),$$

while expressions (82) and (83) for the components of the diffusion thermoeffect tensor can be written in the form

$$v^{(1)} = \frac{kTh^2}{8m_e^2}\left(\frac{3n_e}{\pi}\right)^{2/3}\tau_d\left(\frac{1}{1+\omega^2\tau_d^2}\right.$$
$$+\frac{1}{x_0^2}\frac{2\pi^2}{1+\omega^2\tau_d^2} - \frac{1}{x_0^2}\frac{4\pi^2\omega^2\tau_d^2}{(1+\omega^2\tau_d^2)^2} \qquad (108)$$
$$\left.+\frac{\pi^2}{6}\left(\frac{1}{1+\omega^2\tau_d^2(y^3/x_0^3)}\right)''\bigg|_{x=x_0}\right),$$

$$v^{(2)} = -\frac{kTh^2}{8m_e^2}\left(\frac{3n_e}{\pi}\right)^{2/3}\frac{\omega\tau_d^2}{B}\left(\frac{1}{1+\omega^2\tau_d^2}\right.$$
$$+\frac{33\pi^2}{8x_0^2}\frac{1}{1+\omega^2\tau^2} - \frac{11\pi^2}{2x_0^2}\frac{\omega^2\tau^2}{(1+\omega^2\tau^2)^2} \qquad (109)$$
$$\left.+\frac{\pi^2}{6}\left(\frac{1}{1+\omega^2\tau_d^2(y^3/x_0^3)}\right)''\bigg|_{x=x_0}\right).$$

In Figs. 2 and 3, the ratios of quantities perpendicular and parallel to the magnetic field were calculated for $kT = 0.11E_f$ and $0.2E_f$. The curve corresponding to the phenomenological relation (1) that coincides with the exact solution in the case of strong degeneracy is also shown.

## 5. CONCLUSIONS

In this work, we have found the kinetic tensors of diffusion, thermal diffusion, and diffusion thermoeffect for strongly degenerate nonrelativistic electrons in a nonquantizing magnetic field. A solution is obtained asymptotically exactly in the Lorentz approximation, when electron–electron collisions can be neglected in comparison with electron–nuclei collisions. The tensors are obtained for arbitrary directions of the magnetic field and temperature gradient in Cartesian coordinates according to [8]. In most studies analyzing the kinetic coefficients in astrophysical objects (in particular, the thermal and electric conductivities in neutron stars), the influence of the magnetic field was taken into account phenomenologically by introducing the coefficient $1/(1+\omega^2\tau^2)$, which reduces the heat flux and diffusion across the magnetic field [2, 3]. Our results, obtained by solving the Boltzmann equation, show that the influence of the magnetic field on the kinetic coefficients is stronger and has a more complicated character.

Calculations were carried out for nonrelativistic electrons, although relativistic effects become important in deep layers of the neutron star crust. The main relativistic effect stemming from an increased effective electron mass can be approximately taken into account by replacing the electron rest mass $m_e$ by the relativistic electron mass $m_{e*} = (m_e^2 + p_{Fe}^2/c^2)^{1/2}$ [3]. The vector of diffusion determining the electric current in a medium exhibiting gradients of different parameters in the presence of a nonzero electric field is important for calculating the geometry and evolution of the magnetic field in degenerate stars. The obtained expressions for the kinetic coefficients can be used to calculate the heat fluxes and electric current in white dwarfs, on the surfaces and in the crusts of neutron stars, and in magnetized plasma falling onto a neutron star.

## ACKNOWLEDGMENTS

This work was supported by the Russian Science Foundation, project no. 18-12-00378.